# BINDING ENERGY AND STABILITY OF HEAVY AND SUPERHEAVY NUCLEI

N. N. Kolesnikov*

*Faculty of Physics, Moscow State University, 119991 Moscow, Russia*

Three different ways for description of binding energy of superheavy nuclei are discussed. First, one can consider superheavy nuclei as a part of a whole system of nuclei for which a global mass formula is found. Another way is the detailed local description of energy of superheavy nuclei taking into account the effects of shells and subshells. The third way of description, applied for nuclei in the region limited by principal magic numbers, is attached to the beta-stability line.

## 1 Global optimal mass formula

The problem of superheavy nuclei is typical for all many-particle nuclear system for which the consistent solution with realistic NN-potential is actually unattainable and so different approximate methods (for instance, Hartree–Fock) and simplified potentials (such as Skyrm) are used, or else one address to nuclear models. For the study of superheavy nuclei mostly macro-micro approach is used [1–4] and some mass formulas were proposed that combine the liquid-drop ideology with the shell-model corrections of Strutinsky or Mayers–Swiatecki [5–7]. In order to improve the agreement with experiment the different corrections were introduced in the mass formula [8–10] but then it loses its original simple physical sense and it appears the question whether such modified formula and especially its macroscopic part reflects adequately the requirements of experiment. Because of this in [11] an inverse problem of search for mass formula was considered, independently of any special model, proceeding from the requirement that the minimal rms deviation ($\sigma$) from experimental masses was assured for all nuclei beginning from Ca, but so that it was achieved by introduction of minimal number (n) of parameters. According to [11–13] the global optimal formula for the binding energy $B(A, Z)$ may be written (in MeV)

---

*E-mail: nkoles@mail.ru

as:

$$B(A, Z) = 14.6646A - 31.22A^{1/3} - (0.673 + 0.00029A)Z^2/A^{1/3} \quad (1)$$
$$-(13.164 + 0.004225A)(A - 2Z)^2/A - (3.46 - 0.0928A)|A - 2Z|$$
$$+P(A) + S(Z, N).$$

In Eq. (1) $S(Z, N)$ is the simplest (two-parameter) shell correction and $P(A)$ is the parity correction. At 9 free parameters formula (1) assures the description of all nuclear masses (beginning from $A = 40$) with rms deviation 1.07 MeV, and it describes decently (at least in comparison with other mass formulas) energies of nuclei far from beta-stability line. Note that the nonlinear term $31.22A^{1/3}$ in Eq. (1) differs from the standard expression for surface energy and it corresponds to the relaxation of the surface tension with growth of nuclear sizes. Without introduction of any additional terms or change of parameterizations chosen in [11] Eq. (1) describes sufficiently well the energy of alpha-decay of superheavy nuclei. It is seen in Fig. 1 for nuclei with Z even and in Fig. 2 for Z odd. For transfermium nuclei the better accuracy of Eq. (1) (with $\sigma$ about 0.4 MeV) is achieved if one takes into account the existence of the subshell $N = 170$ [13].

## 2 Beta-stability line and energy of superheavy nuclei

There is however another more effective way for description of heavy and supeheavy nuclei. It was proposed in [13, 14] and is attached to beta stability line, $Z^* = f(A)$, where $Z^*$ is the charge (generally fictitious) of the nucleus with lower mass among all nuclei with fixed mass number A. According to analysis of experiments ( for instance [14, 15]) and in conformity with many-particle shell model [16–19], the charge $Z^*$ of the most stable isobar increases with mass number (A), and in the region limited from above and from below by major magic numbers of neutrons and protons, the dependence of $Z^*$ on A is expressed by a linear function. In particular, for heavy nuclei ($Z > 82, N > 126$)(see [14, 15]):

$$Z^*(A) = 0.356A + 9.1. \quad (2)$$

Besides the energy of $\beta^\pm$ decay $Q_{\beta^\pm}$ of the nucleus $(A, Z)$ depends on its distance from beta-stability line as:

$$Q_{\beta^\pm}(A, Z) = k(Z - Z^*) + D, \quad (3)$$

( where $k$ and $D$ are constants [13, 14]. The confrontation of results of calculation according to Eqs. (2) and (3) with experiment [20] is presented in Fig. 3. For convenience only nuclei with even $Z$ are chosen. In Eq. (3) $k = 1.13$ MeV, the parity correction $D = 0.75$ MeV for even(Z)-odd(N) nuclei and $D = 1.9$ MeV for even(Z)-even(N) nuclei [13, 14]. In view of relations $Q_{\beta^+}(A, Z) = -Q_{\beta^-}(A, Z - 1)$ and $Q_{\beta^-}(A, Z) = Q_{\beta^+}(A, Z + 1)$ the plot in Fig. 3 comprises nuclei of all parities. It is seen that despite of the use in sum only 5 parameters in Eqs. (2) and (3) a sufficiently accurate description for $Q_{\beta^\pm}$ is achieved for all nuclei presented in the table of isotopes [20], the rms deviation is about 0.3 MeV and maximal deviation is less than 0.6 MeV. According to Eqs. (2) and (3) one can calculate the energy of $\beta^\pm$ decay for any heavy or superheavy nucleus knowing only its charge $Z$ and mass number $A$. Moreover (see below) one can calculate also the energies of alpha decay.

In the region $Z > 82, N > 126$ the beta-stability line (see Eq. (2)) deviates from the line $A - 2Z = $ const, corresponding to $\alpha$ decay. The nucleus formed as a result of the $\alpha$ decay of the nucleus $(A, Z)$ approaches to the beta-stability line. For the latter $\Delta Z^*/\Delta A = 0.356$ and for the line of $\alpha$ decay $\Delta Z/\Delta A = 0.5$, so the nucleus $(A - 4, Z - 2)$ approaches closer to the beta-stability line by $\Delta Z = (0.5 - 0.356) \times 4 = 0.576$. Then, as the energy of beta decay is given by Eq. (3), the energy of $\beta$ decay of the nucleus $(A - 4, Z - 2)$ becomes lower than that for nucleus $(A, Z)$ by $0.576k = 0.65$ MeV. At last from the $(\alpha, \beta)$ cycle including nuclei $(A, Z)$, $(A, Z + 1)$, $(A - 4, Z - 2)$ and $(A - 4, Z - 1)$ it follows that the alpha decay energy of the isobar $(A, Z + 1)$ must be higher than that of $(A, Z)$ by 0.65 MeV [13] and this remains valid independently of parity of nucleus. This conclusion is very well confirmed by all (about 200) experimental data for energy of $\alpha$ decay $(Q_\alpha)$, compiled in the table of isotopes [20], the rms deviation is 0.12 MeV at maximal deviation 0.25 MeV, see Table 1.

Following from this, the energy of $\alpha$ decay of the nucleus $(A, Z)$ may be calculated according to the formula

$$Q_\alpha(A, Z) = Q_\alpha^*(A) + 0.65(Z - Z^*), \qquad (4)$$

where $Q_\alpha^*$ is the reduced energy of $\alpha$ decay of the isobaric nucleus (fictitious) lying on the line of beta-stability. The dependence $Q_\alpha^*$ on $A$ characterizes the overall stability of heavy nuclei to the alpha decay. It is shown in Fig. 4 for heavy nuclei with $A > 210$. As it is seen from Fig. 4 the energy of the alpha decay $Q_\alpha^*$ decreases in the beginning up to $A = 236$ but then it increases by steps. For each of such step $Q_\alpha^*$ is well approximated by a linear function of $A$: $Q_\alpha^*(A) = aA+b$. In particular, for the region $212 < A < 224$: $a = -0.275, b = 67.75$; for $224 < A < 234$: $a = -0.1625, b = 42.75$; for $234 < A < 260$: $a = 0.0979, b = -17.99$, and for $A > 260$: $a = 0.0444, b = 4.05$. From this by the use of Eq. (4) one can calculate $Q_\alpha$ for all superheavy nuclei. The results of calculation of $Q_\alpha$ are presented in Table 2. They agree with all experiments [21–23], the rms deviation is 0.2 MeV and maximal deviation is about 0.5 MeV, see Table 3.

It is seen that the approach based on the beta stability line allows to describe the energy of heavy nuclei much better than with optimal mass formula. For instance the rms deviation of $Q_\alpha$ for all heavy nuclei (0.17 MeV) appears one quarter of that for the optimal formula.

## 3   Local description of energy surface

The deviations from regular trends are generally attributed to the effects of shells and subshells. In the region of heavy ($Z > 82$) elements the submagic numbers were attributed (by different authors) to $N = 152, 162, 170$ and to $Z = 100, 114, 104, 108, 110$. Mean while no clear criterion was proposed for their choice. It is natural to relate the (sub)magic numbers to the behaviour of binding energy of neutrons and protons. We proceed from the idea inspired by many- particle shell model [16-19], that in every filling shell the binding energy of neutrons and equally of protons are linear function of $Z$ and $N$.

According to this idea we suppose, that it is possible to divide the whole nuclear energy surface into such domains inside each of which the binding energy of both protons ($B^p$) and neutron ($B^n$) are presentable as a linear functions of number of protons ($Z$) and neutrons ($N$). We shall call such domain limited from all sides by magic numbers as inter-magic regions. Bounding lines between two neighbored regions along $Z = $ const or $N = $ const is reasonable to identify with (sub)magic numbers. Whereas the total binding energy remains continuous always the binding energy of the neutron(proton) can experience a rupture at the cross of (sub)magic number of neutron(proton) or else can change the slope of the line for dependence of binding energy on $N$ or $Z$. According to above idea inside of the intermagic region:

$$B^p_{ij} = p^0_{ij} + a_i(Z - Z_k) + b_i(N - N_l), \qquad (5)$$

where $a_i$, $b_i$ and $p_{ij}^0$ are constants, the index $i$ means the parity of $Z$ and $j$ determines the parity of $N$, $k$ and $l$ are the indices of the region $(k, l)$. Similarly for neutrons $B_{ij}^n$ depends on parameters $c_j$, $d_j$, $p_{ij}^0$ as [10]

$$B_{ij}^n = n_{ij}^0 + c_j(Z - Z_k) + d_j(N - N_l). \qquad (6)$$

The values of parameters $a_i$, $b_i$, $p_{ij}^0$, $c_j$, $d_j$, $n_{ij}^0$ and the (sub)magic numbers themselves were searched by means of solution of an inverse problem at requirement that the experimental binding energy $B^p$ and $B^n$ are reproduced for all heavy and superheavy nuclei compiled in [20] and in [21, 22]. The method of solution is described in [12] and the final results are given in [13]. For convenience the results obtained for $B^p$ and $B^n$ are reduced on the line of beta-stability. In Fig. 5 the reduced energies $B^n$ (see [13]) are presented; the lower curve refer to even($Z$)-even($N$) nuclei; the next line above — to odd($Z$)-even($N$) nuclei; then follows line — of odd-odd nuclei and the last — of even-odd ones; moreover the line denoted as $C$ corresponds to the line averaged over all parities. As it is seen from Fig. 5, after magic number $N = 126$ (fall by 2.1 MeV), the most important subshells are $N = 152$ (fall 0.4 MeV) and $N = 162$ (fall 0.2 MeV). For protons the similar dependence of reduced energy $B^p$ on $Z$ is presented in Fig. 6. It is seen, that for protons after shell $Z = 82$ (fall 1.6 MeV) the most important subshells are $Z = 100$ (fall 0.4 MeV) and $Z = 92$ (fall 0.3 MeV), see in details in [13]. Note that irregularities are often inequal for nuclei of different parity and are connected with change of paring energy. For instance at the cross of the subshell $Z = 100$ the energy $B^p$ decreases by 1 MeV for $Z$ odd, whereas it increases by 0.2 MeV for $Z$ even. And in parallel with this falls the pairing energy of protons.

The subshells correction approach is in fact the method of detailed (local) description of energy surface. It allows to feel the fine peculiarities of energy surface but has need of introduction of a great number of parameters. So, according to [13], for description of energy $Q_\alpha$ for heavy nuclei ($Z > 82$) with accuracy $\sigma = 0.08$ MeV (that is close to the mean experimental error) it is necessary to introduce about 130 parameters. Then the value of the product $\sigma n$ (that characterize the quality of theoretical description) turns out much more than for the case of beta stability line. Moreover without introduction of new parameters the approach of subshell corrections fails in prediction of properties of superheavy nuclei far from region of known studied nuclei. and in this sense it is similar to the approach of Garvay and Kelson [24, 25]. Therefore the most effective way for description of heavy and superheavy nuclei is the use of the line of beta stability.

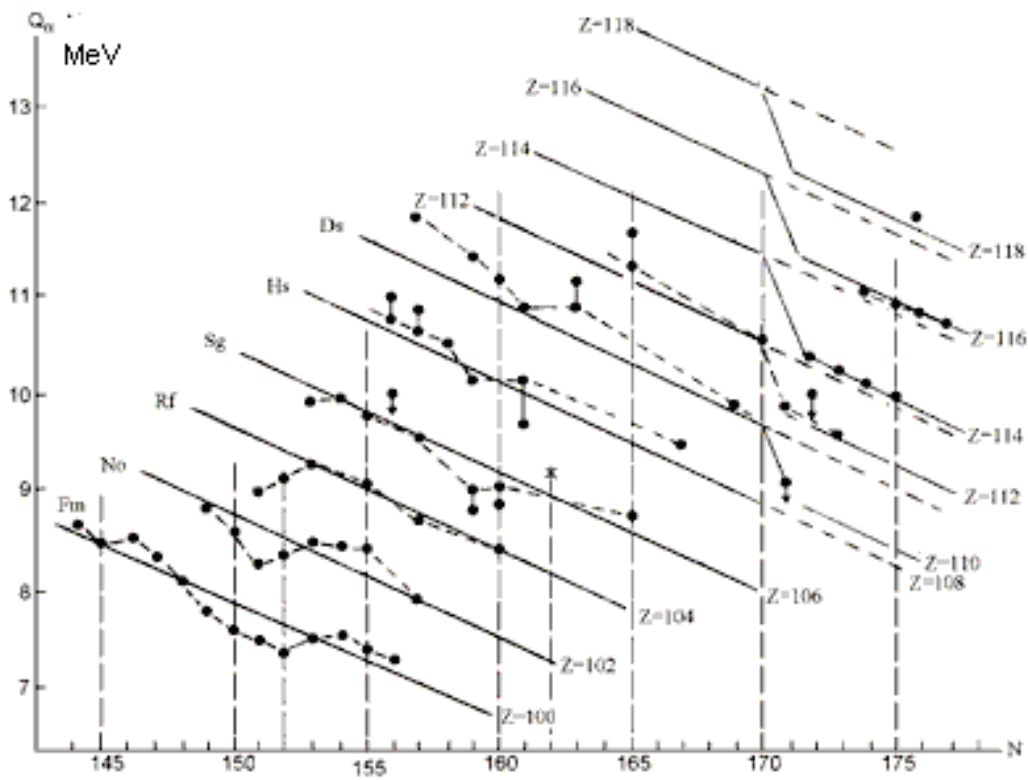

Fig. 1. Energy of alpha decay of transfermium elements.
Comparison with experiment for Z even.

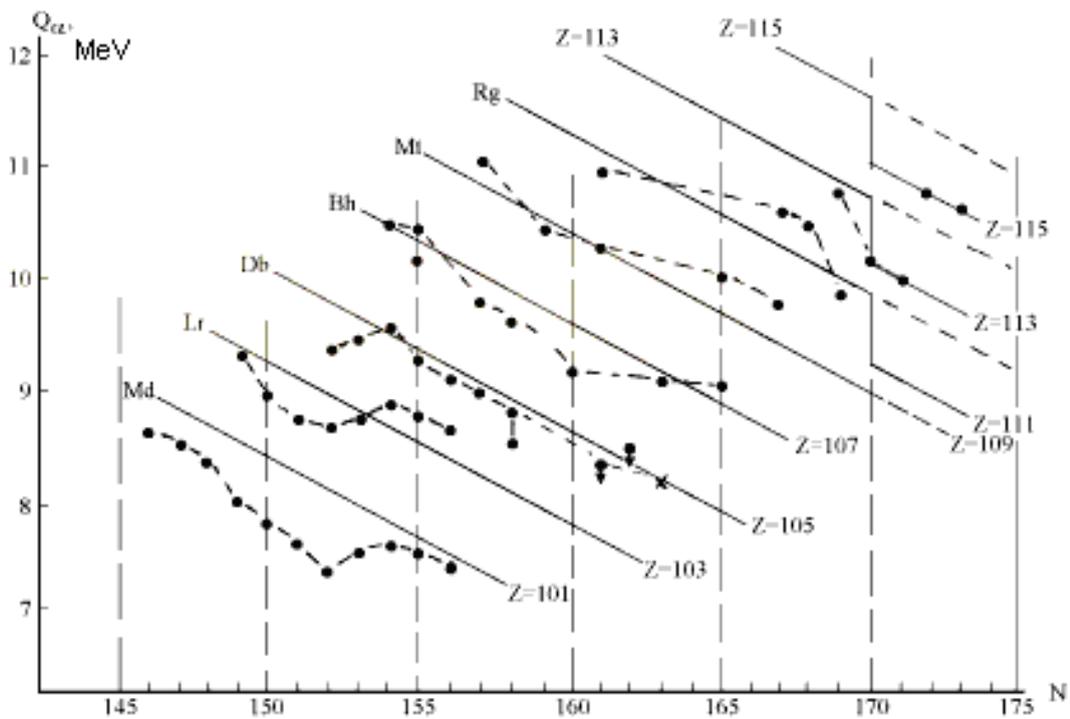

Fig. 2. Energy of alpha decay. Comparison with experiment for Z odd.

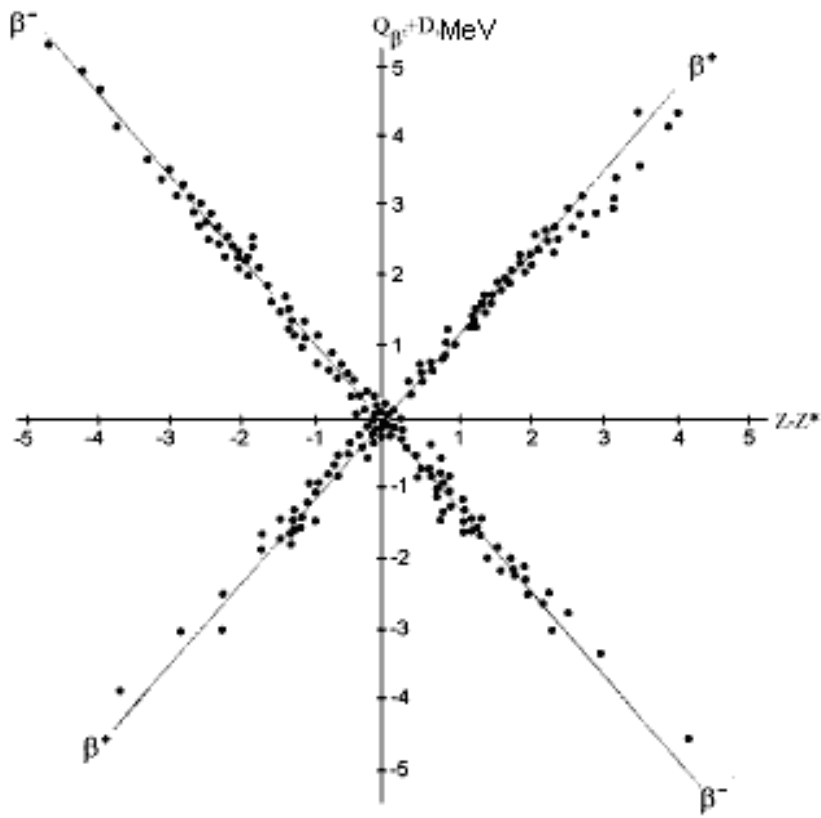

Fig. 3. Dependence of $Q_\beta$ on $Z - Z^*$.

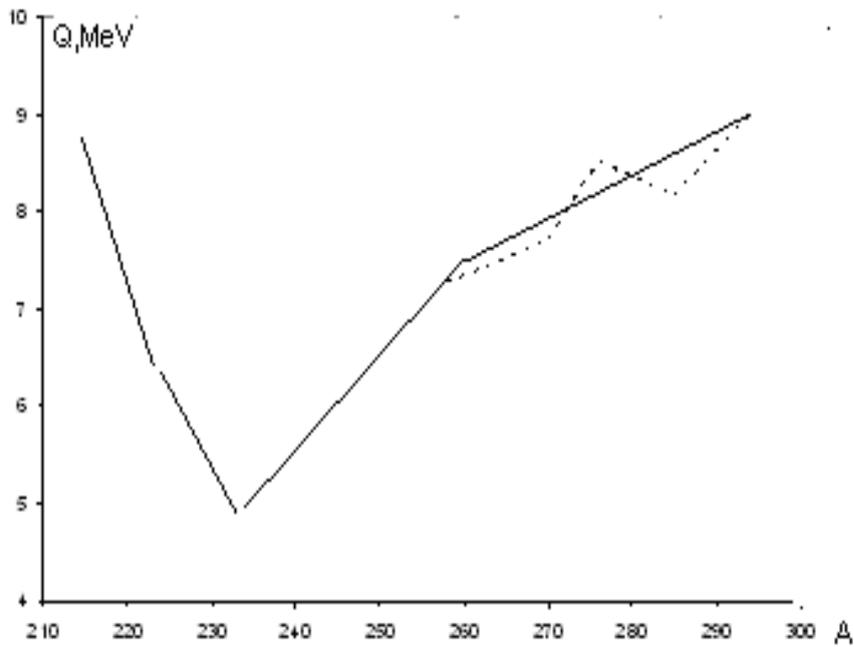

Fig. 4. Alpha stability of heavy nuclei. $Q^*_\alpha$ as a function of A.

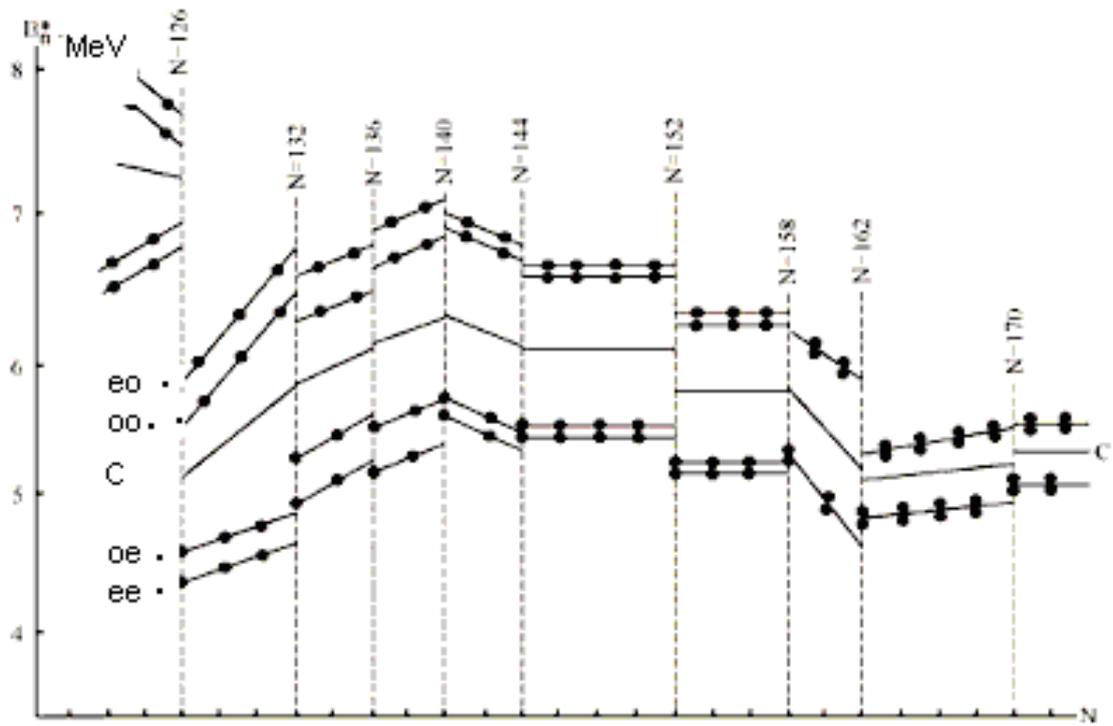

Fig. 5. Reduced binding energy of neutron. Neutron shell effects.

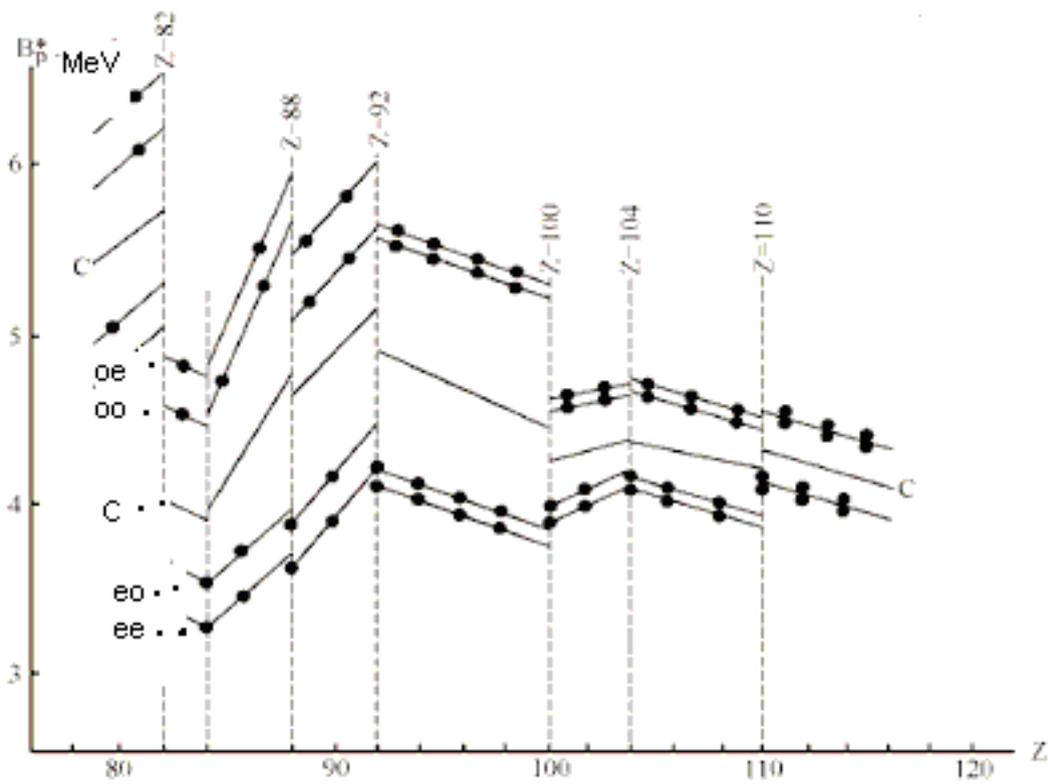

Fig. 6. Reduced binding energy of protons. Proton shell effects.

Table 1. Comparison with experiment [20] the energy of alpha decay $Q_\alpha$ calculated according to Eqs. (2) and (3).

| Z\A | 215 | 216 | 217 | 218 | 219 | 220 |
|---|---|---|---|---|---|---|
| 84 | 7.53  7.55 | 6.91  7.05 | 6.68  6.58 | 6.11  6.06 | (5.50)  5.60 | (5.50)  5.67 |
| 85 | 8.18  8.20 | 7.95  7.70 | 7.20  7.23 | 6.88  6.71 | 6.39  6.25 | 6.40  6.32 |
| 86 | 8.84  8.85 | 8.20  8.35 | 7.89  7.88 | 7.27  7.36 | 6.94  6.90 | 6.81  6.97 |
| 87 | 9.54  9.50 | 9.16  9.00 | 8.47  8.53 | 8.01  8.01 | 7.45  7.55 | 7.59  7.62 |
| 88 |  | 9.53  9.65 | 9.16  9.18 | 8.55  8.66 | 8.13  8.20 | 8.35  8.27 |
| 89 |  |  | 9.82  9.83 | 9.38  9.31 | 8.95  8.85 |  |
| 90 |  |  |  | 9.85  9.96 |  |  |

| Z\A | 221 | 222 | 223 | 224 | 225 | 226 |
|---|---|---|---|---|---|---|
| 86 |  | 5.59  5.34 | (4.88)  4.83 |  |  |  |
| 87 | 6.46  6.48 | 5.81  5.99 | 5.43  5.48 | (5.09)  5.17 |  |  |
| 88 | 6.88  7.13 | 6.68  6.64 | 5.98  6.13 | 5.79  5.82 | (5.18)  5.39 | 4.87  4.94 |
| 89 | 7.86  7.78 | 7.14  7.29 | 6.78  6.78 | 6.32  6.47 | 5.94  6.04 | 5.54  5.59 |
| 90 | 8.63  8.43 | 8.13  7.94 | 7.45  7.43 | 7.31  7.12 | 6.92  6.69 | 6.45  6.24 |
| 91 | (9.13)  9.08 | 8.59 | 8.35  8.38 | 7.63  7.77 | 7.38  7.34 | 6.99  6.89 |
| 92 |  | (9.96)  9.24 |  |  |  | 7.56  7.54 |

| Z\A | 227 | 228 | 229 | 230 | 231 | 232 |
|---|---|---|---|---|---|---|
| 88 | (4.36)  4.62 |  |  |  |  |  |
| 89 | 5.04  5.27 | (4.76)  4.91 |  |  | 3.91  3.72 |  |
| 90 | 6.15  5.92 | 5.52  5.56 | 5.17  5.14 | 4.76  4.77 | 4.20  4.37 | 4.08  4.08 |
| 91 | 6.58  6.57 | 6.23  6.21 | 5.84  5.79 | 5.44  5.42 | 5.15  5.02 | 4.61  4.73 |
| 92 | (7.20)  7.22 | 6.80  6.86 | 6.47  6.44 | 5.93  6.07 | 5.55  5.67 | 5.41  5.38 |
| 93 |  | (7.61)  7.51 | 7.01  7.09 | 6.78  6.72 | 6.37  6.32 | (6.00)  6.03 |
| 94 |  |  | (8.11)  7.74 | (7.44)  7.37 |  | 6.72  6.68 |

| Z\A | 233 | 234 | 235 | 236 | 237 | 238 |
|---|---|---|---|---|---|---|
| 91 | 4.34  4.33 |  | 3.98  3.97 |  |  |  |
| 92 | 4.91  4.98 | 4.87  4.83 | 4.68  4.62 | 4.56  4.53 | 4.23  4.32 | 4.27  4.13 |
| 93 | (5.70)  5.63 | 5.36  5.48 | 5.19  5.27 | 5.07  5.18 | 4.96  4.97 | 4.68  4.78 |
| 94 | 6.42  6.28 | 6.27  6.13 | 5.96  5.92 | 5.88  5.83 | 5.75  5.62 | 5.59  5.43 |
| 95 | (7.06)  6.93 | (6.73)  6.78 |  | (6.34)  6.48 | (6.20)  6.27 | 6.04  6.08 |
| 96 |  |  |  |  |  | 6.63  6.73 |
| 97 |  |  |  |  |  | 7.29  7.38 |

| Z\A | 239 | 240 | 241 | 242 | 243 | 244 |
|---|---|---|---|---|---|---|
| 92 | 4.11  3.99 | (3.92) 3.80 | | | | |
| 93 | 4.57  4.64 | 4.25  4.45 | 4.25  4.35 | (4.07) 4.30 | | (3.59) 4.01 |
| 94 | 5.24  5.29 | 5.26  5.10 | 5.14  5.00 | 4.98  4.95 | 4.75  4.85 | 4.67  4.66 |
| 95 | 5.92  5.94 | 5.59  5.75 | 5.64  5.65 | 5.59  5.60 | 5.46  5.50 | 5.24  5.31 |
| 96 | (6.50) 6.59 | 6.40  6.40 | 6.18  6.30 | 6.22  6.25 | 6.17  6.15 | 5.90  5.96 |
| 97 | | (7.27) 7.05 | (7.03) 6.95 | (6.96) 6.90 | 6.87  6.80 | 6.78  6.61 |
| 98 | | | | 7.51  7.55 | (7.40) 7.45 | 7.33  7.26 |
| 99 | | | | (7.84) 8.20 | | |

| Z\A | 245 | 246 | 247 | 248 | 249 | 250 |
|---|---|---|---|---|---|---|
| 94 | (4.40) 4.50 | | | | | |
| 95 | 5.16  5.15 | (4.93) 4.95 | (4.71) 4.69 | (4.61) 4.57 | (4.64) 4.35 | (4.28) 4.31 |
| 96 | 5.62  5.80 | 5.48  5.60 | 5.35  5.34 | 5.16  5.22 | 5.17  5.00 | 5.27  4.96 |
| 97 | 6.45  6.45 | (6.15) 6.25 | 5.89  5.99 | (5.68) 5.87 | 5.53  5.65 | 5.61  5.61 |
| 98 | 7.26  7.10 | 6.87  6.90 | (6.55) 6.64 | 6.57  6.52 | 6.30  6.30 | 6.13  6.26 |
| 99 | 7.86  7.75 | (7.70) 7.55 | 7.44  7.29 | (7.15) 7.17 | 6.88  6.95 | (6.73) 6.91 |
| 100 | (8.40) 8.40 | 8.37  8.20 | (8.20) 7.94 | 8.00  7.82 | | 7.55  7.56 |

| Z\A | 251 | 252 | 253 | 254 |
|---|---|---|---|---|
| 97 | (5.68) 5.43 | | | |
| 98 | 6.18  6.08 | 6.22  6.02 | 6.13  6.10 | 5.93  5.95 |
| 99 | 6.60  6.73 | 6.73  6.67 | 6.74  6.75 | 6.62  6.60 |
| 100 | 7.42  7.38 | 7.15  7.32 | 7.20  7.40 | 7.30  7.25 |
| 101 | (8.05) 8.03 | (7.85) 7.97 | 8.56  8.70 | (7.80) 7.90 |
| 102 | | 8.56  8.62 | | 8.34  8.55 |

For every value of A in the left column the experimental value of $Q_\alpha$ is given (uncertain in brackets) whereas in the right one — the calculated values.

Table 2. Energy $Q_\alpha$ calculated for transfermium elements.

| Z | | | | | | | | | | | | | | | |
|---|---|---|---|---|---|---|---|---|---|---|---|---|---|---|---|
| 112 | | | | | | | | | 11.95 | 12.01 | 11.78 | 11.65 | 11.67 | 11.42 |
| 111 | | | | | | 11.75 | 11.62 | 11.63 | 11.46 | 11.52 | 11.30 | 11.36 | 11.13 | 11.30 | 11.02 |
| 110 | | | | | 11.66 | 11.43 | 11.20 | 10.97 | 10.98 | 10.81 | 10.87 | 10.65 | 10.71 | 10.48 | 10.65 |
| 109 | | | | 11.33 | 11.10 | 11.01 | 10.78 | 10.65 | 10.32 | 10.33 | 10.16 | 10.22 | 10.00 | 10.16 | 9.83 |
| 108 | | | | 10.97 | 10.68 | 10.45 | 10.36 | 10.19 | 9.90 | 9.67 | 9.68 | 9.61 | 9.57 | 9.35 | 9.41 |
| 107 | | | 10.77 | 10.54 | 10.31 | 10.03 | 9.91 | 9.71 | 9.48 | 9.25 | 9.02 | 9.03 | 8.96 | 8.92 | 8.70 |
| 106 | 10.57 | 10.58 | 10.36 | 10.12 | 9.84 | 9.66 | 9.38 | 9.15 | 9.06 | 8.83 | 8.60 | 8.37 | 8.38 | 8.31 | 8.27 |
| 105 | 10.09 | 9.92 | 9.93 | 9.71 | 9.47 | 9.24 | 9.01 | 8.73 | 8.50 | 8.41 | 8.18 | 7.95 | 7.72 | 7.73 | 7.66 |
| 104 | 9.50 | 9.44 | 9.27 | 9.28 | 9.06 | 8.82 | 8.59 | 8.36 | 8.08 | 7.85 | 7.76 | 7.53 | 7.30 | 7.07 | 7.08 |
| 103 | 9.26 | 8.85 | 8.79 | 8.62 | 8.63 | 8.41 | 8.17 | 7.94 | 7.71 | 7.43 | 7.20 | 7.11 | 6.88 | 6.65 | 6.42 |
| 102 | 8.93 | 8.61 | 8.20 | 8.14 | 7.97 | 7.98 | 7.76 | 7.82 | 7.29 | 7.06 | 6.78 | 6.55 | 6.46 | | |
| 101 | 8.03 | 7.68 | 7.96 | 7.55 | 7.49 | 7.32 | 7.33 | 7.11 | 6.87 | 6.64 | 6.41 | 6.13 | | | |
| 100 | 7.52 | 7.42 | 7.08 | 7.31 | 6.90 | 6.84 | 6.67 | 6.68 | 6.46 | 6.22 | | | | | |
| N = | 151 | 152 | 153 | 154 | 155 | 156 | 157 | 158 | 159 | 160 | 161 | 162 | 163 | 164 | 165 |

| Z | | | | | | | | | | | | | | |
|---|---|---|---|---|---|---|---|---|---|---|---|---|---|---|
| 120 | | | | | 13.49 | 13.40 | 13.29 | 13.18 | 13.01 | 12.87 | 12.73 | 12.60 | | |
| 119 | | | | | 12.98 | 12.84 | 12.75 | 12.64 | 12.51 | 12.36 | 12.22 | 12.08 | 11.95 | |
| 118 | | | | | 12.46 | 12.33 | 12.19 | 12.10 | 11.99 | 11.86 | 11.71 | 11.57 | 11.43 | |
| 117 | | | | 12.07 | 11.93 | 11.81 | 11.68 | 11.54 | 11.45 | 11.34 | 11.21 | 11.06 | 10.92 | |
| 116 | | | 11.70 | 11.64 | 11.46 | 11.28 | 11.16 | 11.03 | 10.89 | 10.80 | 10.69 | 10.56 | 10.41 | |
| 115 | | 11.65 | 11.45 | 11.05 | 10.98 | 10.81 | 10.63 | 10.51 | 10.38 | 10.24 | 10.15 | 10.04 | 9.91 | |
| 114 | 11.86 | 11.75 | 11.39 | 11.00 | 10.80 | 10.40 | 10.36 | 10.16 | 9.98 | 9.86 | 9.73 | 9.59 | 9.50 | 9.39 |
| 113 | 11.83 | 11.21 | 11.10 | 10.74 | 10.35 | 10.15 | 9.75 | 9.69 | 9.50 | 9.33 | 9.21 | 9.08 | 8.94 | |
| 112 | 11.36 | 11.18 | 10.56 | 10.45 | 10.09 | 9.70 | 9.50 | 9.10 | 9.04 | 8.85 | 8.68 | 8.56 | | |
| 111 | 10.75 | 10.71 | 10.53 | 9.91 | 9.80 | 9.44 | 9.05 | 8.85 | 8.45 | 8.39 | 8.20 | | | |
| 110 | 10.39 | 10.10 | 10.06 | 9.88 | 9.26 | 9.15 | 8.79 | 8.40 | 8.20 | 7.80 | | | | |
| 109 | 10.08 | 9.74 | 9.45 | 9.41 | 9.23 | 8.61 | 8.50 | 8.14 | 7.75 | | | | | |
| 108 | 9.48 | 9.35 | 9.09 | .8.80 | 8.76 | 8.58 | 7.96 | | | | | | | |
| 107 | 9.10 | 8.83 | 8.70 | 8.44 | 8.15 | 8.11 | | | | | | | | |
| 106 | 8.52 | 8.45 | 8.18 | 8.05 | 7.79 | | | | | | | | | |
| N = | 166 | 167 | 168 | 169 | 170 | 171 | 172 | 173 | 174 | 175 | 176 | 177 | 178 | 179 |

Table 3. Comparison with experiment calculated values of $Q_\alpha$
for transfermium elements [21, 22, 23].

| Nucleus | | | Exp. | Calc. | Nucleus | | | Exp. | Calc. |
|---|---|---|---|---|---|---|---|---|---|
| A | Z | N | MeV | | A | Z | N | MeV | |
| 294 | 117 | 177 | 10.96(10) | 11.21 | 266 | 106 | 161 | S.F. | 8.60 |
| 290 | 115 | 175 | 10.09(40) | 10.58 | 294 | 118 | 176 | 11.81(6) | 11.86 |
| 286 | 113 | 173 | 9.77(10) | 9.53 | 290 | 116 | 174 | 10.80(7) | 11.03 |
| 282 | 111 | 171 | 9.14(10) | 9.28 | 286 | 114 | 172 | 10.33(6) | 10.36 |
| 278 | 109 | 169 | 9.68(19) | 9.25 | 282 | 112 | 170 | ≤ 10.69 | 10.03 |
| 274 | 107 | 167 | 8.94(10) | 8.67 | 293 | 116 | 177 | 10.69(6) | 10.69 |
| 270 | 105 | 165 | S.F. | 7.66 | 289 | 114 | 175 | 9.98(5) | 9.86 |
| 293 | 117 | 176 | 11.17(8) | 11.34 | 285 | 112 | 173 | 9.28(5) | 9.10 |
| 289 | 115 | 174 | 10.45(9) | 10.51 | 281 | 110 | 171 | ≤9.10 | 9.15 |
| 254 | 113 | 171 | 9.88(8) | 10.15 | 291 | 116 | 175 | 10.89(7) | 10.89 |
| 288 | 115 | 173 | 10.61(6) | 10.68 | 287 | 114 | 173 | 10.16(6) | 10.16 |
| 284 | 113 | 171 | 10.13(6) | 10.15 | 283 | 112 | 171 | 9.62(6) | 9.70 |
| 280 | 111 | 169 | 9.87(6) | 9.91 | 279 | 110 | 169 | 9.84(6) | 9.88 |
| 276 | 109 | 167 | 9.85(6) | 9.74 | 275 | 108 | 167 | 9.44(6) | 9.36 |
| 272 | 107 | 165 | 9.19(6) | 8.70 | 271 | 106 | 165 | 8.67(8) | 8.27 |
| 268 | 105 | 163 | S.F. | 7.72 | 288 | 114 | 174 | 10.08(6) | 9.98 |
| 287 | 115 | 172 | 10.74(9) | 10.63 | 284 | 112 | 172 | ≤9.81 | 9.50 |
| 283 | 113 | 170 | 10.26(9) | 10.35 | 290 | 116 | 174 | 11.00(8) | 11.03 |
| 279 | 111 | 168 | 10.52(10) | 10.53 | 286 | 114 | 172 | 10.33(6) | 10.36 |
| 274 | 109 | 166 | 10.48(9) | 10.08 | 282 | 113 | 169 | 10.78(8) | 10.74 |
| 282 | 113 | 169 | 10.78(8) | 10.74 | 270 | 107 | 163 | 9.06(8) | 9.96 |
| 278 | 111 | 167 | 10.85(8) | 10.71 | | | | | |